\documentclass[journal]{IEEEtran}

\usepackage{cite}
\usepackage{graphicx}
\usepackage[cmex10]{amsmath}
\usepackage{amsfonts,amssymb,amsthm}
\usepackage{mathrsfs}
\usepackage{enumerate}
\usepackage[makeroom]{cancel}
\usepackage{color}
\usepackage[utf8]{inputenc}
\usepackage[english]{babel}
\usepackage[percentage]{overpic}

\usepackage[font=footnotesize]{caption}
\usepackage{subfig}
\usepackage{subfloat}

\theoremstyle{remark}

\begin{document}

\title{\huge{Non-Orthogonal Multiple Access combined with\\Random Linear Network Coded Cooperation}}
\author{Amjad~Saeed~Khan,~\IEEEmembership{Student Member,~IEEE,} and Ioannis~Chatzigeorgiou,~\IEEEmembership{Senior Member,~IEEE}
	\thanks{A. S. Khan and I. Chatzigeorgiou are with the School of Computing and Communications, Lancaster University, Lancaster, United Kingdom (e-mail: \{a.khan9, i.chatzigeorgiou\}@lancaster.ac.uk).}
}

\maketitle

\begin{abstract}
This letter considers two groups of source nodes. Each group transmits packets to its own designated destination node over single-hop links and via a cluster of relay nodes shared by both groups. In an effort to boost reliability without sacrificing throughput, a scheme is proposed, whereby packets at the relay nodes are combined using two methods; packets delivered by different groups are mixed using non-orthogonal multiple access principles, while packets originating from the same group are mixed using random linear network coding. An analytical framework that characterizes the performance of the proposed scheme is developed, compared to simulation results and benchmarked against a counterpart scheme that is based on orthogonal multiple access.
\end{abstract}

\begin{IEEEkeywords}
Network coding, non-orthogonal multiple access, sparse random matrices, decoding probability, throughput.
\end{IEEEkeywords}

%
%

\section{Introduction}

Random Linear Network Coding (RLNC) is a scheme that allows an intermediate node to combine and forward the data of multiple users in a single transmission, and can effectively improve  network capacity~\cite{Multicast_capacity_2}. RLNC has the inherent capability to achieve spatial diversity. For example, it has been shown in~\cite{NC_diversity} that network coding can improve the diversity gain of networks that either contain distributed antenna systems or support cooperative relaying. Furthermore, RLNC can improve both the throughput~\cite{Multicast_capacity_2} and the latency in a network~\cite{RLNC_latency} by reducing the number of distinct transmissions. 

The benefits of network coding have made it an attractive solution for challenges encountered in existing and future communication systems. For instance, it has been shown in~\cite{OFDM-NC} that by modifying the IEEE 802.11g frame structure, network coding combined with Orthogonal Frequency Division Multiplexing (OFDM) can significantly improve throughput. The importance of network-coded cooperation has been demonstrated in~\cite{NCC_tutorial} and implemented in~\cite{OFDMA-NCC} with Orthogonal Frequency Multiple Access (OFDMA). Recently, Non-Orthogonal Multiple Access (NOMA) has been recognised as a promising multiple access technique for 5G mobile networks~\cite{NOMA_sig_letter, NOMA_5G}. It has been shown in~\cite{NOMA-OFDM},~\cite{NOMA-OFDM2} that combining NOMA with OFDM can improve the spectral efficiency and accommodate more users than the conventional OFDMA-based systems. Moreover, the usefulness of RLNC for downlink NOMA-based transmissions has been studied in~\cite{RLNC_NOMA}.

This letter considers network-coded cooperation in a NOMA-based scenario with two groups of source nodes. Each group communicates with a different destination node via multiple relay nodes. To the best of our knowledge, this work represents the first attempt to characterise the performance of NOMA-based RLNC cooperation. The main contributions of our work can be summarized as follows: 
(i) we propose a framework which integrates the benefits of NOMA-based multiplexing and RLNC-based cooperative relaying;
(ii) using the fundamentals of RLNC and uplink/downlink NOMA, we derive closed-form expressions for the network performance, in terms of the decoding probability at each node, and the system throughput;
(iii) we validate the accuracy of the derived expressions through simulations and we investigate the impact of the system parameters on the network performance and throughput.

%
%

\section{System Model}
\label{sec:sys_mod}

Consider a network with two source groups, two destination nodes and $N$ commonly shared relay nodes $\mathrm{r}_1,\mathrm{r}_2,\hdots, \mathrm{r}_N$. Each source group $\mathrm{G}_k$ contains $K$ source nodes $\mathrm{s}_1^{(k)},\mathrm{s}_2^{(k)},\hdots, \mathrm{s}_K^{(k)}$ for $k=1,2$. The packets transmitted by source nodes in $\mathrm{G}_k$ are meant to be received by destination $\mathrm{d}_k$, either directly or via relay nodes. The acceptable transmission rate for $\mathrm{G}_1$ is $R_1^{*}$ and for $\mathrm{G}_2$ is $R_2^{*}$. Without loss of generality, we assume that all source nodes in $\mathrm{G}_1$ require a comparatively high quality of service with $R_1^{*}<R_2^{*}$. In practice, $\mathrm{G}_1$ could be a group of devices (e.g., sensors) associated to high risk applications that need to be connected quickly with low data rate, and $\mathrm{G}_2$ could be a group of devices related to low risk applications that can afford opportunistic connectivity. All nodes operate in half duplex mode. The links connecting the nodes are modeled as quasi-static Rayleigh fading channels. The channel gain between nodes $i$ and $j$ is represented by $h_{ij}$, which is a zero-mean circularly symmetric complex Gaussian random variable with variance $\sigma_{ij}^2$. 

Before the communication process is initiated, source nodes from the two groups are paired according to their indices, such that $\mathrm{s}_i^{(1)}$ in group $\mathrm{G}_1$ is paired with $\mathrm{s}_i^{(2)}$ in $\mathrm{G}_2$. Only paired nodes are allowed to transmit simultaneously over the same frequency band. The simultaneous transmission of two nodes exploits the principle of superposition coding, which is a key component of NOMA. Node pairing in NOMA has been recently proposed for 3GPP Long Term Evolution Advanced (LTE-A)~\cite{user_pairing}. Source nodes in different pairs transmit over orthogonal frequency bands, and therefore can be recovered independently. This approach is also known as OFDM-NOMA~\cite{NOMA-OFDM} but, for the sake of brevity, we shall simply refer it to as NOMA. We consider the worst case scenario, in which both source groups contain an equal (i.e., $K$) number of source nodes, such that relay nodes always receive superimposed signals. The proposed communication process is divided into the broadcast phase and  the relay phase. 

During the \textit{broadcast phase}, each source node broadcasts a packet in the form of an information-bearing signal to the relay and destination nodes. The signals transmitted by the $i^{th}$ source pair $\bigl(\mathrm{s}_i^{(1)},\mathrm{s}_i^{(2)}\bigr)$, and received by a relay node $\mathrm{r}_j$ and destination nodes $\mathrm{d}_1$ and $\mathrm{d}_2$, are respectively given by
\begin{equation}
\label{eq:rel_des_broadcast}
\begin{split}
z_{\mathrm{r}_j}^{i}&=\sqrt{\alpha_1P_\mathrm{s}}h_{\mathrm{s}_i^{(1)}\mathrm{r}_j}{\tilde{x}}_{i}+\sqrt{\alpha_2P_\mathrm{s}}h_{\mathrm{s}_i^{(2)}\mathrm{r}_j}\tilde{y}_{i}+w_{\mathrm{r}_j}^i,\notag\\
z_{\mathrm{d}_1}^{i}&=\sqrt{\alpha_1P_\mathrm{s}}h_{\mathrm{s}_i^{(1)}\mathrm{d}_1}\tilde{x}_i+w_{\mathrm{d}_1}^i,\\
z_{\mathrm{d}_2}^{i}&=\sqrt{\alpha_2P_\mathrm{s}}h_{\mathrm{s}_i^{(2)}\mathrm{d}_2}\tilde{y}_i+w_{\mathrm{d}_2}^i,\notag
\end{split}
\end{equation}
where $P_\mathrm{s}$ is the total transmission power by the source pair, $\alpha_1$ and $\alpha_2$ are the fractions of $P_\mathrm{s}$ transmitted by $\mathrm{s}_i^{(1)}$ and $\mathrm{s}_i^{(2)}$, respectively, with $\alpha_1+\alpha_2=1$, and $\{\tilde{x}_i, \tilde{y}_i\}$ represent the modulated signals of data packets $\{x_{i}, y_{i}\}$. 
The additive white Gaussian noise components at the relay and destination nodes are represented by $w_{\mathrm{r}_j}^i$ and $w_{\mathrm{d}_k}^i$, respectively. All relay nodes employ Successive Interference Cancellation (SIC) to recover the transmitted signals, and then disjointly demodulate and store the correctly received packets. 

During the \textit{relay phase}, a relay node $\mathrm{r}_j$ employs RLNC on the successfully received and stored data packets of groups $\mathrm{G}_1$ and $\mathrm{G}_2$, and generates coded packets $m_{j}^{(1)}$ and $m_{j}^{(2)}$, respectively, given by $m_{j}^{(1)}=\sum_{i=1}^{K}c_{i,j}^{(1)}x_{i}$ and $m_{j}^{(2)}=\sum_{i=1}^{K}{c}_{i,j}^{(2)}y_{i}$, where, $c_{i,j}^{(k)}$ represents the coding coefficients over a finite field $F_q$ of size $q$. The value of a coefficient is zero if a received packet contains irrecoverable errors; otherwise, the value of that coefficient is selected uniformly at random from the remaining $q-1$ elements of $F_q$. The probability mass function of $c_{i,j}^{(k)}$ is given as 
\begin{equation}
\mathrm{Pr}\bigl(c_{i,j}^{(k)}=t\bigr)
=\left\{%
\begin{IEEEeqnarraybox}[\relax][c]{l's}
\epsilon_{\mathrm{s}_i^{(k)}\mathrm{r}_j},&for $t=0$,\\%
\frac{1-\epsilon_{\mathrm{s}_i^{(k)}\mathrm{r}_j}}{q-1},&for $t\in F_q\setminus\{0\}$,%
\end{IEEEeqnarraybox}\right.
\end{equation}
where $0\leq\epsilon_{\mathrm{s}_i^{(k)}\mathrm{r}_j}\leq 1$ is the outage probability of the link connecting the source node $\mathrm{s}_i^{(k)}$with the relay node $\mathrm{r}_j$. The closed form expression of $\epsilon_{\mathrm{s}_i^{(k)}\mathrm{r}_j}$will be presented in Section~\ref{sec:link_outage}. This type of RLNC at the relay nodes is known as $\textit{sparse}$ RLNC, where the sparsity level is determined by the outage probability $\epsilon_{\mathrm{s}_i^{(k)}\mathrm{r}_j}$~\cite{Blomer_1997},~\cite{Improve_bounds}.

Each relay node, instead of transmitting two separate network-coded signals (one for each destination), generates a signal that is the superposition of the two network-coded signals and broadcasts it to both destinations. Relay transmissions are orthogonal, either in time or in frequency. The superimposed signal transmitted by relay $\mathrm{r}_j$ can be expressed as $(\sqrt{P_\mathrm{r}\beta_{1}}\tilde{m}_{j}^{(1)}$+$\sqrt{P_\mathrm{r}\beta_{2}}\tilde{m}_{j}^{(2)})$, where $P_\mathrm{r}$ is the total transmitted power, and  $\beta_{1}$, $\beta_{2}$ denote the power allocation coefficients, such that $\beta_{1}+\beta_{2}=1$ with $\beta_1>\beta_2$ in order to satisfy the quality of service requirement~\cite{QOS_requirement}. Thus, the received signal at destination $\mathrm{d}_k$ is given as
\begin{equation}
\hat{z}_{\mathrm{d}_k}^j=h_{\mathrm{r}_j\mathrm{d}_k}(\sqrt{P_\mathrm{r}\beta_{1}}\tilde{m}_{j}^{(1)}+\sqrt{P_\mathrm{r}\beta_{2}}\tilde{m}_{j}^{(2)})+\hat{w}_{\mathrm{d}_k}^j\notag
\end{equation}
where $\hat{w}_{\mathrm{d}_k}^j$ is the Gaussian noise component. Each destination node employs SIC in order to separate the superimposed signals and retrieve the relevant coded packets. Destination $\mathrm{d}_k$ will recover the data packets of source group $\mathrm{G}_k$ if it collects $K$ linearly independent packets directly from that source group and via the relay nodes.

%
%

\section{Achievable rate and link outage probability} 
\label{sec:link_outage}

This section describes the achievable transmission rate of source-to-destination, source-to-relay and relay-to-destination links. An outage occurs when the achievable rate is less than the target rate of transmission. Therefore, the outage probability of each link can be expressed in terms of the corresponding achievable rate and the target rate. 

Let us first consider the broadcast phase, during which signals arrive at each destination node directly from the respective source group. The achievable rate of the ${{\mathrm{s}_i^{(k)}\mathrm{d}_k}}$ link, which originates from group $\mathrm{G}_k$, can be obtained as
\begin{equation}
\label{eq:ref_5}
R_{\mathrm{s}_i^{(k)}\mathrm{d}_k}=B_\mathrm{s}\log\big(1+ \frac{P_\mathrm{s} \alpha_k|h_{\mathrm{s}_i^{(k)}\mathrm{d}_k}|^2}{B_sN_0}\big)
\end{equation}
where $k\in\{1,2\}$, $i\in\{1,2,\hdots,K\}$, $N_0$ represents the noise power and $B_s$ denotes the bandwidth of the frequency band allocated to each source pair for simultaneous transmissions, as discussed in Section~\ref{sec:sys_mod}. The outage probability of the ${{\mathrm{s}_i^{(k)}\mathrm{d}_k}}$ link can be derived if we combine expression \eqref{eq:ref_5} with the cumulative distribution function of Rayleigh fading \cite[eq.~(7.6)]{Goldsmith2005}, which gives
\begin{equation}
\epsilon_{\mathrm{s}_i^{(k)}\mathrm{d}_k}=\mathrm{Pr}(R_{\mathrm{s}_i^{(k)}\mathrm{d}_k}\leq R_k^{*})=1-\exp(-\frac{\tau_k}{\rho_\mathrm{s} \alpha_k\sigma_{\mathrm{s}_i^{(k)}\mathrm{d}_k}^2})\notag
\end{equation}
where $\rho_\mathrm{s}=\frac{P_\mathrm{s}}{B_\mathrm{s} N_0}$ and $\tau_k=2^{R_k^{*}/B_\mathrm{s}}-1$. The achievable rate of the link between one of the nodes of a source pair and a relay node $\mathrm{r}_j$ depends on the channel conditions of both links that connect the nodes of the source pair with $\mathrm{r}_j$. For example, assume that $\alpha_1|h_{\mathrm{s}_i^{(1)}\mathrm{r}_j}|>\alpha_2|h_{\mathrm{s}_i^{(2)}\mathrm{r}_j}|$. In that case, SIC at the relay node $\mathrm{r}_j$ will first recover the signal of the node from $\mathrm{G}_1$ and treat the other signal as interference. Thus, the achievable rate of a link between $\mathrm{s}_i^{(k)}$ and $\mathrm{r}_j$ can be expressed as~\cite{NOMA_rate_ref}
\begin{equation}
R_{\mathrm{s}_i^{(1)}\mathrm{r}_j}=B_\mathrm{s}\log\big(1+\frac{\alpha_1|h_{\mathrm{s}_i^{(1)}\mathrm{r}_j}|^2}{\alpha_2|h_{\mathrm{s}_i^{(2)}\mathrm{r}_j}|^2+1/\rho_\mathrm{s}}\big)
\end{equation}
\begin{equation}
R_{\mathrm{s}_i^{(2)}\mathrm{r}_j}=B_\mathrm{s}\log\big(1+\rho_\mathrm{s} \alpha_2|h_{\mathrm{s}_i^{(2)}\mathrm{r}_j}|^2\big).
\end{equation}
The outage probability of a link between $\mathrm{s}_i^{(k)}$ and $\mathrm{r}_j$ can be obtained as $\epsilon_{\mathrm{s}_i^{(1)}\mathrm{r}_j}=\mathrm{Pr}(R_{\mathrm{s}_i^{(k)}\mathrm{r}_j}<R_{k}^{*})$, thus
\begin{equation}
\epsilon_{\mathrm{s}_i^{(1)}\mathrm{r}_j}=1-\frac{\alpha_1\sigma_{\mathrm{s}_i^{(1)}\mathrm{r}_j}^2}{\tau_1\alpha_2\sigma_{\mathrm{s}_i^{(2)}\mathrm{r}_j}^2+\alpha_1\sigma_{\mathrm{s}_i^{(1)}\mathrm{r}_j}^2}\exp(-\frac{\tau_1}{\rho_\mathrm{s} \alpha_1\sigma_{\mathrm{s}_i^{(1)}\mathrm{r}_j}^2})\notag
\end{equation}
\begin{equation}
\resizebox{0.4866\textwidth}{!}{$
\begin{split}
&\epsilon_{\mathrm{s}_i^{(2)}\mathrm{r}_j}=1-\mathrm{Pr}\bigl[(R_{\mathrm{s}_i^{(1)}\mathrm{r}_j}>R_1^{*})\cap(R_{\mathrm{s}_i^{(2)}\mathrm{r}_j}>R_2^{*})\bigr]\\
&=1-\frac{\alpha_1\sigma_{\mathrm{s}_i^{(1)}\mathrm{r}_j}^2}{\tau_1\alpha_2\sigma_{\mathrm{s}_i^{(2)}\mathrm{r}_j}^2+\alpha_1\sigma_{\mathrm{s}_i^{(1)}\mathrm{r}_j}^2}\exp(-\frac{\tau_1(\tau_2+1)}{\rho_\mathrm{s} \alpha_1\sigma_{\mathrm{s}_i^{(1)}\mathrm{r}_j}^2}-
\frac{\tau_2}{\rho_\mathrm{s} \alpha_2\sigma_{\mathrm{s}_i^{(2)}\mathrm{r}_j}^2}).\notag
\end{split}$}
\end{equation}

During the relay phase, the destination node $\mathrm{d}_2$ can only successfully recover the coded signals corresponding to source group $\mathrm{G}_2$, when $R_{\mathrm{r}_j\mathrm{d}_2}\!>\!R_2^{*}$ provided that 
$R_{\mathrm{r}_j\mathrm{d}_1}>R_1^{*}$. On the other hand, the destination $\mathrm{d}_1$ can recover the coded signals of $\mathrm{G}_1$, when $R_{\mathrm{r}_j\mathrm{d}_1}>R_1^{*}$. The achievable rates are given as
\begin{equation}
R_{\mathrm{r}_j\mathrm{d}_1}=B_\mathrm{s}\log\big(1+\frac{\beta_1|h_{\mathrm{r}_j\mathrm{d}_1}|^2}{\beta_2|h_{\mathrm{r}_j\mathrm{d}_1}|^2+1/\rho_\mathrm{r}}\big)
\end{equation}
\begin{equation}
R_{\mathrm{r}_j\mathrm{d}_2}=B_\mathrm{s}\log\big(1+\rho_\mathrm{r} \beta_2|h_{\mathrm{r}_j\mathrm{d}_2}|^2\big)
\end{equation}
where $B_\mathrm{s}$ is the bandwidth allocated to each pair of relays, and $\rho_\mathrm{r}=\frac{P_\mathrm{r}}{B_\mathrm{s}N_0}$. It is assumed that $\beta_1 \geq \tau_1 \beta_2$, otherwise the outage probability is always one~\cite{NOMA_sig_letter}. The outage probability of links ${{\mathrm{r}_j\mathrm{d}_1}}$ and ${{\mathrm{r}_j\mathrm{d}_2}}$ can be respectively obtained as
\begin{equation}
\begin{split}
\epsilon_{\mathrm{r}_j\mathrm{d}_1}&=\mathrm{Pr}(\frac{\beta_1|h_{\mathrm{r}_j\mathrm{d}_1}|^2}{\beta_2|h_{\mathrm{r}_j\mathrm{d}_1}|^2+1/\rho_\mathrm{r}}\leq \tau_1)\\
&=1-\exp(-\frac{\tau_1}{(\rho_\mathrm{r}\beta_1-\tau_1\rho_\mathrm{r}\beta_2)\sigma_{\mathrm{r}_j\mathrm{d}_1}^2}),\notag
\end{split}
\end{equation}
\begin{equation}
\begin{split}
\epsilon_{\mathrm{r}_j\mathrm{d}_2}&=1-\mathrm{Pr}(\frac{\beta_1|h_{\mathrm{r}_j\mathrm{d}_2}|^2}{\beta_2|h_{\mathrm{r}_j\mathrm{d}_2}|^2+1/\rho_\mathrm{r}}> \tau_1,\rho_\mathrm{r} \beta_2|h_{\mathrm{r}_j\mathrm{d}_2}|^2>\tau_2)\\
&=1-\exp(-\frac{1}{\rho_\mathrm{r} \sigma_{\mathrm{r}_j\mathrm{d}_2}^2}\max(\frac{\tau_1}{\beta_1-\tau_1\beta_2},\frac{\tau_2}{\beta_2})).\notag
\end{split}
\end{equation}
\subsection*{OMA-based Benchmark scheme}
In this letter, we consider conventional OFDMA as the benchmark Orthogonal Multiple Access (OMA) scheme. According to this scheme, all nodes $\mathrm{s}_i^{(k)}$ and $\mathrm{r}_j$ transmit over orthogonal frequency bands. As a result, likewise~\eqref{eq:ref_5}, the achievable rates of source-to-relay and source-to-destination links during the broadcast phase, and the relay-to-destination links during the relay phase can be respectively obtained as
\begin{equation}
\begin{split}
R_{\mathrm{s}_i^{(k)}u}&=\frac{B_\mathrm{s}}{2}\log(1+ \frac{P_\mathrm{s} \alpha_k|h_{\mathrm{s}_i^{(k)}u}|^2}{0.5B_\mathrm{s}N_0}),\notag\\
R_{\mathrm{r}_j \mathrm{d}_k}&=\frac{B_\mathrm{s}}{2}\log(1+ \frac{P_r \beta_k|h_{\mathrm{r}_j\mathrm{d}_k}|^2}{0.5B_\mathrm{s}N_0})\notag
\end{split}
\end{equation}
where $u\in\{\mathrm{r}_j,\mathrm{d}_k\}$. The factor $1/2$ is due to the fact that, unlike NOMA, each sub-band is now further split between two transmitting nodes. Note that, using the achievable rates, we can derive the outage probabilities. These results can be further extended to RLNC-based analysis, which will be presented in the next section, and can be used as benchmarks against the proposed NOMA-based scheme.

In the remainder of the letter, we assume that links connecting co-located transmitting nodes with receiving nodes are statistically similar, hence $\epsilon_{\mathrm{r}_j\mathrm{d}_k}\!=\!\epsilon_{\mathrm{r} \mathrm{d}_k}$, $\epsilon_{\mathrm{s}_i^{(k)}\mathrm{r}_j}\!=\!\epsilon_{\mathrm{s}^{(k)}\mathrm{r}}$ and $\epsilon_{\mathrm{s}_i^{(k)}\mathrm{d}_k}\!=\!\epsilon_{\mathrm{s}^{(k)}\mathrm{d}_k}$ for all valid values of $i$, $j$ and $k$.

%
%
 
\section{Decoding probability and Analysis}
\label{sec:dec_prob_analysis}
This section analyses the system performance in terms of the probability of a destination node successfully recovering the packets of all nodes in the corresponding source group. Furthermore, the system throughput is derived as a function of the number of packet transmissions.

The destination node $\mathrm{d}_k$ can recover the packets of all source nodes in group $\mathrm{G}_k$ if and only if it collects packets that yield $K$ degrees of freedoms (dofs). Note that dofs at a destination node represent successfully received linearly independent packets, which can be either source packets delivered during the broadcast phase, or coded packets transmitted during the relay phase. According to \cite[eq.~(5)]{Improve_bounds} and \cite[eq.~(8)]{Improve_bounds}, the probability that the $N\geq K$ coded packets, which have been transmitted by the $N$ relay nodes, will yield $K$ dofs can be bounded as follows:
\begin{IEEEeqnarray}{lCr}
\displaystyle{P^{\prime}(K, N,\epsilon_{\mathrm{s}^{(k)}\mathrm{r}},q)\!\geq\max\Big\{\prod_{i=1}^{K}\!\bigl(1-\Gamma_{\max}^{N-i+1}\bigr),1\!-\!\!\sum_{w=1}^{K}\!\binom{K}{w}}\times\nonumber\\
\times(q-1)^{w-1}\big[q^{-1}+(1-q^{-1})\big(1-\frac{1-\epsilon_{\mathrm{s}^{(k)}\mathrm{r}}}{1-q^{-1}}\big)^w\big]^{N}\Big\}\label{eq:dep_prb}
\end{IEEEeqnarray}
where $\Gamma_{\max}=\max\big\{\epsilon_{\mathrm{s}^{(k)}\mathrm{r}},\displaystyle\frac{1-\epsilon_{\mathrm{s}^{(k)}\mathrm{r}}}{q-1}\big\}$. 

In order to formulate the decoding probability at each destination node, let us assume that the destination $\mathrm{d}_k$ successfully received $m$ packets, given that $K+N$ packets were transmitted, i.e., $K$ source packets during the broadcast phase and $N$ coded packets during the relay phase. If we denote by $f_{\ell}(N_\mathrm{T},\epsilon)$ the probability mass function of the binomial distribution, that is,
\begin{equation} 
f_{\ell}(N_\mathrm{T},\epsilon)={\binom{N_\mathrm{T}}{\ell}{\epsilon}^{N_\mathrm{T}-\ell}(1-\epsilon)^\ell}
\end{equation}
then the probability that $h$ of the $m$ packets are source packets and the remaining $m-h$ are coded packets is given by
\begin{equation} 
\label{eq:rec_prb}
P_{h/m}(\epsilon_{\mathrm{s}^{(k)}\mathrm{d}_k},\epsilon_{\mathrm{r}\mathrm{d}_k})=f_h(K,\epsilon_{\mathrm{s}^{(k)}\mathrm{d}_k})f_{m-h}(N,\epsilon_{\mathrm{r} \mathrm{d}_k}).
\end{equation}

The contribution of the $h$ recovered source packets to the $m-h$ coded packets can be removed, so that the $m-h$ coded packets become linear combinations of the remaining $K-h$ source packets only. Thus, at this point of the decoding process, the destination node $\mathrm{d}_k$ can successfully recover the remaining data packets if and only if the modified $m-h$ coded packets yield $K-h$ dofs. By employing \eqref{eq:dep_prb}, \eqref{eq:rec_prb} and the law of total probability, the overall decoding probability at the destination $\mathrm{d}_k$ can be expressed as
\begin{equation}
\resizebox{0.4866\textwidth}{!}{$
P_\mathrm{\mathrm{d}_k}(K,N)=\displaystyle\!\!\sum_{m=K}^{N+K}\sum_{h=h_\mathrm{min}}^{K}\!\!\!P_{h/m}(\epsilon_{\mathrm{s}^{(k)}\mathrm{d}_k},\epsilon_{\mathrm{r}\mathrm{d}_k})P^{\prime}(K-h,m-h,\epsilon_{\mathrm{s}^{(k)}\mathrm{r}},q)
\label{eq:overall_dec_prob}
$}
\end{equation}
where $h_\mathrm{\min}=\max(0,m-N)$. 

Note that retransmissions are not allowed in case of packet failures during the broadcast phase or the relay phase. Therefore, by modifying the expression of the end-to-end throughput in~\cite{throughput_formula}, the average system throughput can be defined as
\begin{equation}
\eta=\frac{K}{K+\max\{E_{\mathrm{d}_1}(N),E_{\mathrm{d}_2}(N)\}}
\end{equation}
where $E_{\mathrm{d}_k}(N)$ is the average number of relay nodes needed by each destination node $\mathrm{d}_k$ to recover the entire source group $\mathrm{G}_k$, and can be calculated using~\cite{decoding_delay}
\begin{equation}
\label{eq:avg_delay}
E_{\mathrm{d}_k}(N)=N-\sum_{v=0}^{N-1}P_\mathrm{\mathrm{d}_k}(K,v).
\end{equation}
Moreover, by following \eqref{eq:avg_delay}, the average number of relays required for both destinations to decode the packets of the respective source groups can be represented as $E_\mathrm{T}(N)=N-\sum_{v=0}^{N-1}P_\mathrm{joint}(K,v)$, where \mbox{$P_\mathrm{joint}(K,v)=P_\mathrm{\mathrm{d}_1}(K,v)P_\mathrm{\mathrm{d}_2}(K,v)$}.

\begin{figure*}[t]
\captionsetup[subfigure]{labelformat=empty}
\subfloat[Figure~1: Simulation results and performance comparison between NOMA-RLNC and OMA-RLNC, when $K=20$, $N=10$ and $q=4$.]{\label{fig:fig_1}\includegraphics[width=5.8cm, height=4.3cm]{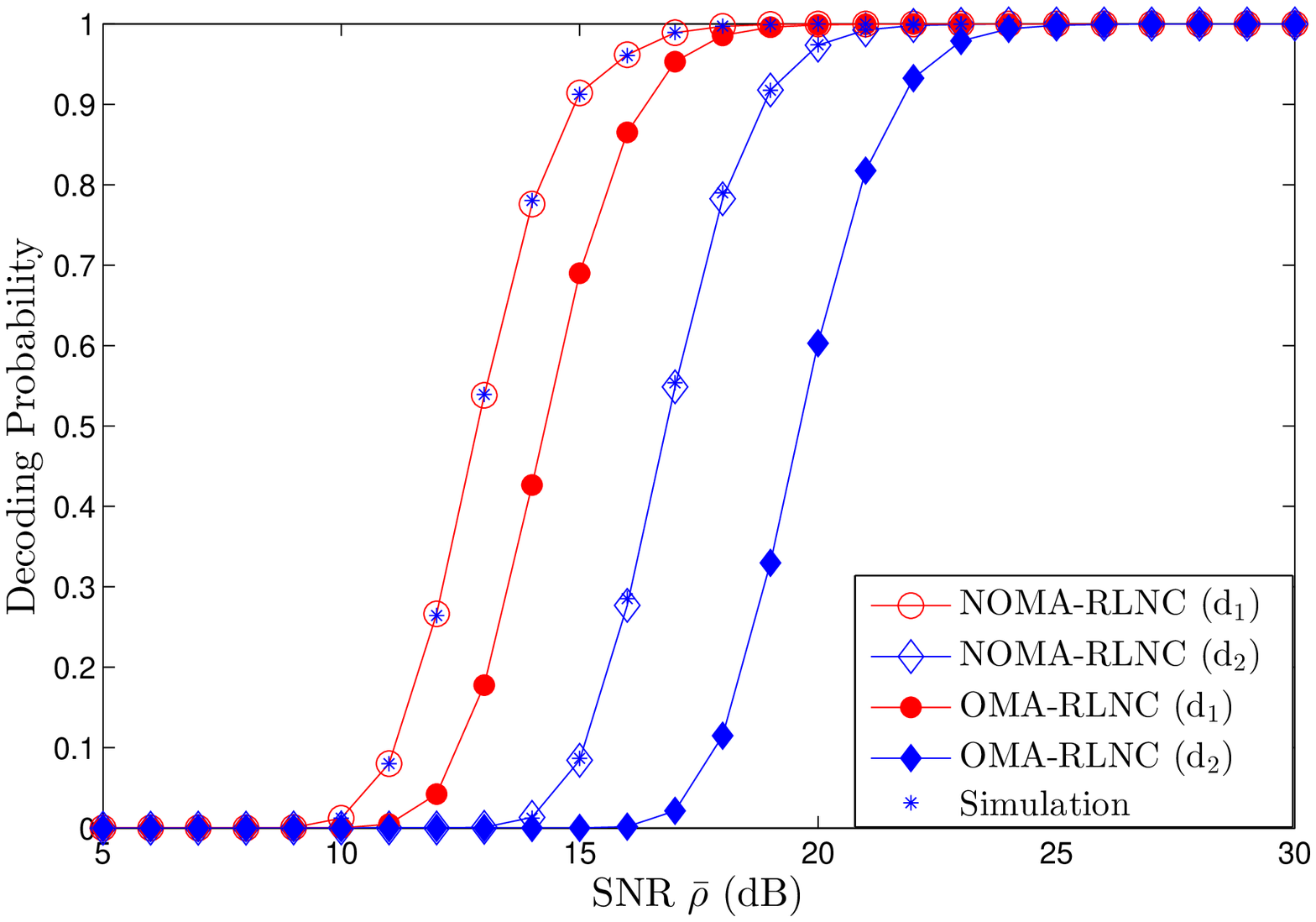}}
\:\:\:\subfloat[Figure~2: Effect of the field size $q$ and the number of relay nodes $N$ on the joint decoding probability, when $K=20$.]{\includegraphics[width=5.8cm, height=4.3cm]{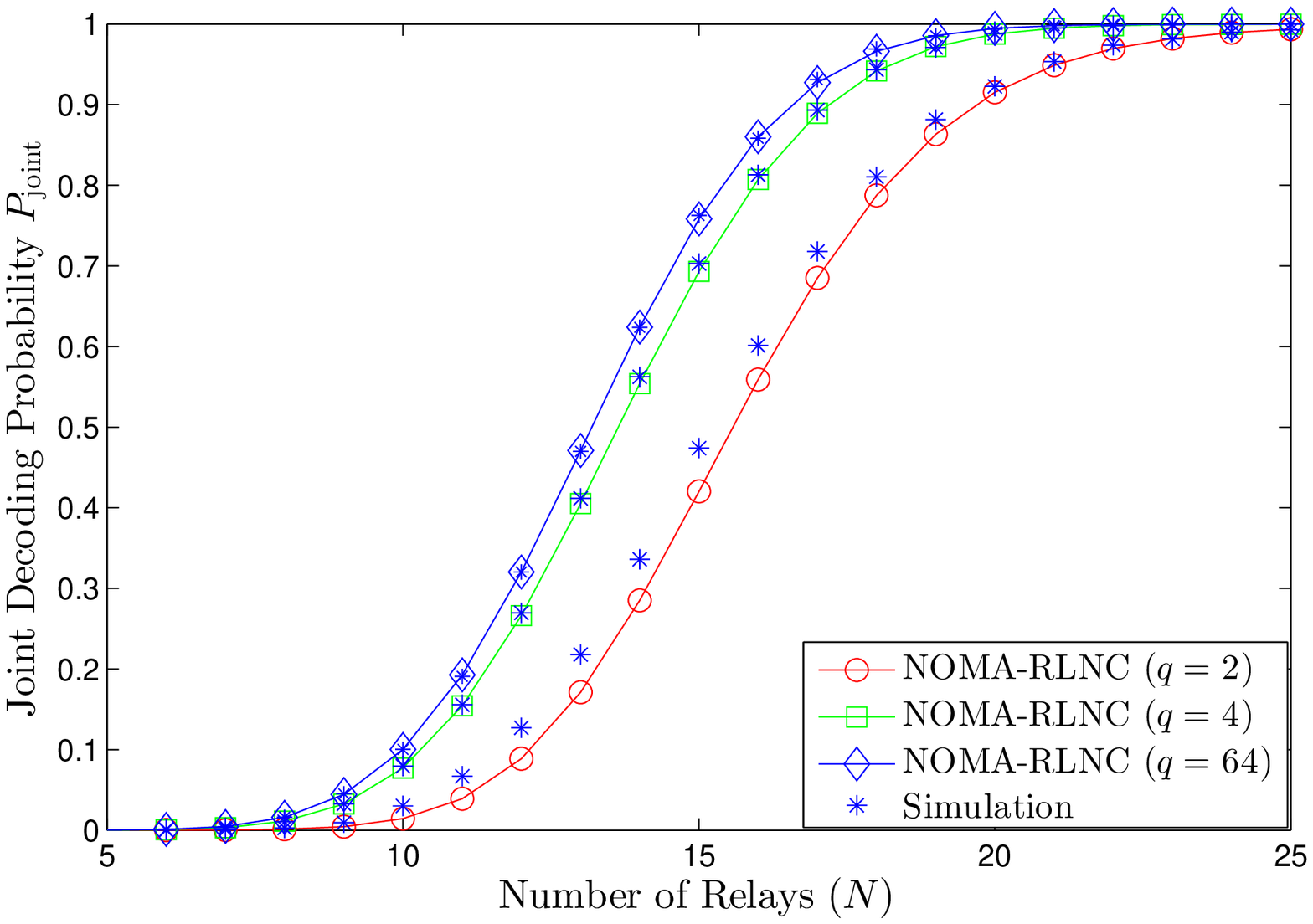}}\label{fig:fig_2}
\:\:\:\subfloat[Figure~3: Comparison between the two schemes in terms of the required average number of relay nodes and the SNR when $K=20$ and $q=4$.]{\includegraphics[width=5.8cm, height=4.3cm]{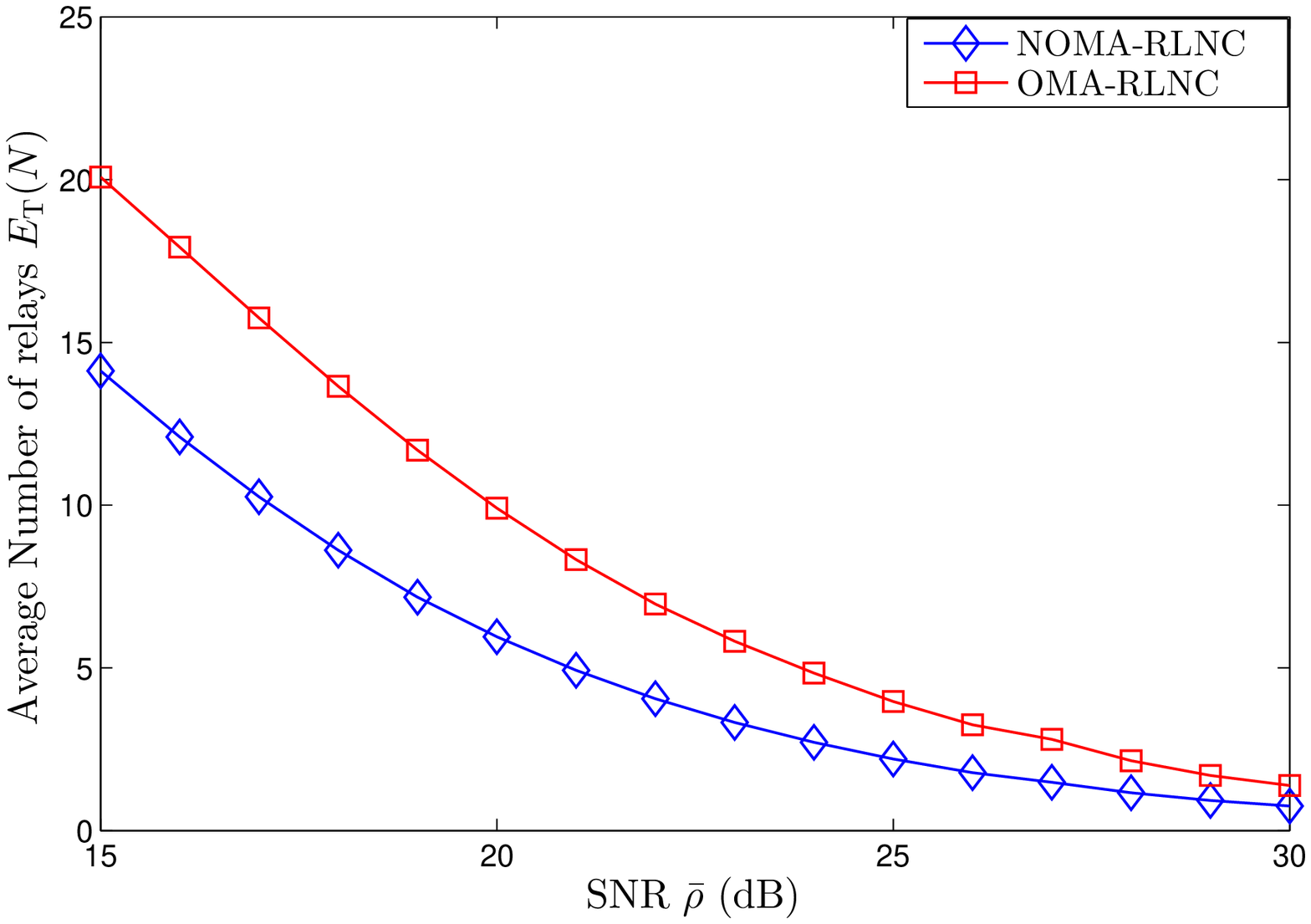}}\label{fig:fig_3}
\vspace{-4mm}
\end{figure*}

%
%

\section{Numerical Results}

In this section, the accuracy of the derived analytical bound in \eqref{eq:dep_prb}, when used in combination with the decoding probability in \eqref{eq:overall_dec_prob}, is verified through simulations. In the considered system setup, the bandwidth of each sub-band is normalized to 1, i.e., \mbox{$B_\mathrm{s}=1$}. The source nodes and relay nodes have been positioned such that $\sigma_{\mathrm{s}^{(1)}\mathrm{d}_1}^2=0.1458$, $\sigma_{\mathrm{s}^{(2)}\mathrm{d}_2}^2=0.1458$, $\sigma_{\mathrm{s}^{(1)}\mathrm{r}}^2=2.9155$, $\sigma_{\mathrm{s}^{(2)}\mathrm{r}}^2=1$, $\sigma_{\mathrm{r}\mathrm{d}_1}^2=1.3717$ and $\sigma_{\mathrm{r}\mathrm{d}_2}^2 =1.9531$. We set $\alpha_1=0.6$ and $\alpha_2=0.4$, while exhaustive search has been used to identify the values of $\beta_1$ and $\beta_2$ that maximize the joint decoding probability mentioned in Section~\ref{sec:dec_prob_analysis}. The average system SNR is set equal to $\rho_\mathrm{s}=\rho_\mathrm{r}=\bar{\rho}$ and, unless otherwise stated, we consider $R_1^*=1$, $R_2^*=1.5$.

Fig.~1 shows the decoding probabilities $P_\mathrm{d_1}$ and $P_\mathrm{d_2}$ at the two destination nodes in terms of the system SNR. The figure clearly demonstrates the tightness of the analytical curve to the simulation results. The decoding probability $P_\mathrm{d_1}$ is greater than $P_\mathrm{d_2}$ because node $\mathrm{d}_1$ supports a lower target rate than node $\mathrm{d}_2$, and $\mathrm{d}_1$ is allocated more power than $\mathrm{d}_2$ to ensure that the quality of service requirements are met. As expected, NOMA-RLNC outperforms OMA-RLNC because each source node in NOMA-RLNC benefits from being allocated twice the bandwidth that is allocated in OMA-RLNC.

Fig.~2 shows the joint decoding probability, for different values of field size $q$, as a function of the number of relays. The analytical bound is close to the simulation results for $q=2$ and becomes tighter for greater values of $q$. A significant gain in performance can be observed when the field size increases from $q=2$ to $q=4$. However, the increase in gain is markedly smaller when $q$ further increases from $4$ to $64$. This is because the certainty of linear independence between coded packets increases with the field size and approaches the highest possible degree even for relatively small values of $q$. We stress that the computational complexity of the decoder at the destination nodes also depends on the value of $q$. Thus, the choice of the field size over which RLNC is performed results in a trade-off between complexity and performance gain.     

Fig.~3 illustrates the relationship between the system SNR and the average number of relays required for the decoding of the source packets of both source groups by the respective destination nodes. The curves establish the diversity advantage offered by the combination of NOMA with RLNC as opposed to OMA with RLNC. For a fixed value of SNR, OMA-RLNC clearly needs more relays for cooperation than NOMA-RLNC. Alternatively, OMA-RLNC can achieve the same performance as NOMA-RLNC at the expense of a higher SNR.    

Fig.~4 presents the system throughput as a function of the system SNR, for different target rates. The performance gap between NOMA-RLNC and OMA-RLNC is evident. We observe that, for a fixed SNR value, when the target rate increases from $R^{*}_2=1.5$ to $R^{*}_2=2$, the outage probability increases and, therefore, the system throughput reduces. Interestingly, an increase in the target rate also increases the performance gap between NOMA-RLNC and OMA-RNC, that is, the throughput degradation of NOMA-RLNC is less severe than that of OMA-RLNC. An intuitive reason for this observation is that the $1/2$ spectral loss in OMA dominates the system throughput. 

\begin{figure}[h]
\captionsetup{labelformat=empty}
\centering
\includegraphics[width=0.8\columnwidth]{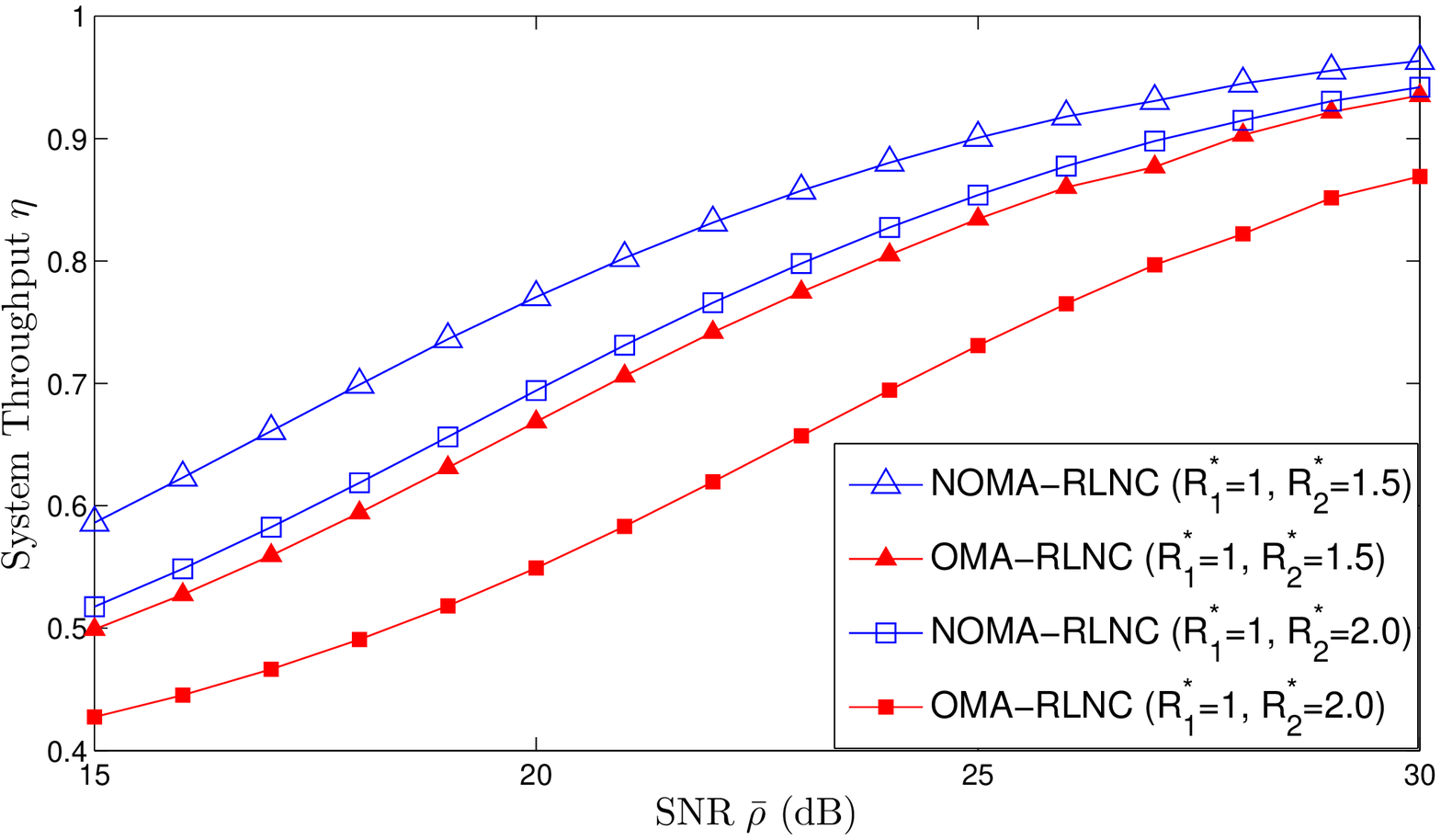}
\caption{Figure 4: Effect of target rates on the system throughput against the system SNR, when $K=20$ and $q=4$.}
\vspace{-4.9mm}
\label{fig:fig_4}
\end{figure}

%
%

\section{Conclusions}

This letter investigated the benefits of NOMA-based multiplexing and RLNC-based cooperative relaying in terms of decoding probability and system throughput. Simulation results established the tightness of the derived expressions. Comparisons emphasized the importance of network-coded cooperation and demonstrated the impact of the filed size on network performance. This work showed that the combination of NOMA with RLNC can clearly provide a superior performance, in terms of diversity gain and system throughput, than the combination of conventional OMA with RLNC. We note that network performance can be further improved if modern techniques of node pairing \cite{NOMA_pairing} are employed.

\bibliographystyle{IEEEtran}
\bibliography{IEEEabrv,IEEE_letter_Ref}
\end{document}